\documentclass[conference]{IEEEtran}
\IEEEoverridecommandlockouts

\usepackage{amsmath,amssymb,amsfonts}
\usepackage{placeins}
\usepackage[utf8]{inputenc}
\usepackage{graphicx}
\usepackage{epstopdf}
\usepackage{booktabs}
\usepackage{textcomp}
\usepackage{bm}
\usepackage{tcolorbox}
\usepackage{braket}
\usepackage{amsthm}
\usepackage[textsize=tiny]{todonotes}
\usepackage{yquant}
\usepackage{hyperref}
\usepackage{mathtools}

\usepackage{circuitikz}
\usepackage{tikz}
\usetikzlibrary{angles, arrows.meta}
\usepackage{tkz-euclide}
\usepackage{float}
\usepackage{caption}
\usepackage{subcaption}
\usepackage{tabularx}
\usepackage{csvsimple}
\usepackage{xcolor}
\usepackage{comment}

\usepackage{gensymb}
\usepackage{multirow}
\usepackage{svg}
\usepackage{algorithm} 
\usepackage{algpseudocode} 
\usepackage{csquotes}
\usepackage{array}
\usepackage{adjustbox}

\usetikzlibrary{shapes.geometric, positioning, quantikz}

\usepackage[font=footnotesize]{subcaption}
\usepackage{yquant}
\useyquantlanguage{groups}
\usepackage{pgfplots}

\newtheorem{example}{Example}

\newcolumntype{R}{>{\raggedleft\arraybackslash}p{0.8cm}}

\newcolumntype{?}{!{\vrule width 1.5pt}}

\usepackage{circuitikz}
\usetikzlibrary{fit}

\usepackage{xcolor}
\usepackage{colortbl}
\usepackage{pgfplots} 
\usepackage{array}

\newcommand{\cellcolorval}[1]{%
    \ifnum#1<0
        \pgfmathsetmacro{\value}{100*(#1+75)/75}
        \edef\temp{\noexpand\cellcolor{white!\value!green}}%
        \temp #1%
    \else
        \pgfmathsetmacro{\value}{100*(#1)/75} 
        \edef\temp{\noexpand\cellcolor{red!\value!white}}%
        \temp #1%
    \fi
}

\usepackage[style=ieee, maxnames=4, minnames=1, date=year, doi=false,isbn=false,backend=biber]{biblatex}

\usetikzlibrary{calc}

\def\BibTeX{{\rm B\kern-.05em{\sc i\kern-.025em b}\kern-.08em
    T\kern-.1667em\lower.7ex\hbox{E}\kern-.125emX}}

\DeclareFieldFormat
  [misc, book]
  {title}{\mkbibquote{#1}}

\AtEveryBibitem{
  \clearname{editor}%
  \clearfield{series}%
  \clearfield{isbn}%
  \clearfield{issn}%
  \clearfield{volume}%
  \clearfield{number}%
  \clearfield{pages}%
  \clearfield{url}%
  \clearfield{doi}%
}

\def\authorwidth{0.7cm}
\begin{document}
\newtheorem{exmp}{Example}

\title{Quantum Circuit Optimization for the \\Fault-Tolerance Era: \\Do We Have to Start from Scratch?}

\author{
	\IEEEauthorblockN{ Tobias V. Forster\IEEEauthorrefmark{1}\hspace*{\authorwidth}
Nils Quetschlich\IEEEauthorrefmark{1}\hspace*{\authorwidth}
Robert Wille\IEEEauthorrefmark{1}\IEEEauthorrefmark{2}\IEEEauthorrefmark{3}}
\IEEEauthorblockA{\IEEEauthorrefmark{1}Chair for Design Automation, Technical University of Munich, Germany}
\IEEEauthorblockA{\IEEEauthorrefmark{2}Munich Quantum Software Company GmbH, Garching near Munich, Germany}
 \IEEEauthorblockA{\IEEEauthorrefmark{3}Software Competence Center Hagenberg GmbH (SCCH), Austria}
\IEEEauthorblockA{\{t.forster, nils.quetschlich,  robert.wille\}@tum.de 
\hspace*{\authorwidth}
\url{https://www.cda.cit.tum.de/research/quantum}}
\vspace{-5mm}
}

\maketitle

\begin{abstract}
Quantum computing has made significant advancements in the last years in both hardware and software.
Unfortunately, the currently available \mbox{\emph{Noisy Intermediate-Scale Quantum} (NISQ)} hardware is still heavily affected by noise. 
Many optimization techniques have been developed to reduce the negative effects thereof, which, however, only works up to a certain point.
Therefore, scaling quantum applications from currently considered small research examples to industrial applications requires \mbox{error-correction} techniques to execute quantum circuits in a \mbox{fault-tolerant} fashion and enter the \mbox{\emph{Fault-Tolerant Quantum Computing} (FTQC)} era.
These \mbox{error-correction} techniques introduce dramatic qubit overheads, leading to the requirement of tens of thousands of qubits already for toy-sized examples.
Hence, quantum circuit optimization that reduces qubit overheads when shifting from the NISQ to the FTQC era is essential.
This raises the question, whether we need to start from scratch, or whether current \mbox{state-of-the-art} optimization techniques can be used as a basis for this.
To approach this question, this work investigates the effects of different optimization passes on a representative selection of quantum circuits.
Since hardly any tools to automatically design and evaluate FTQC quantum circuits exist yet, we utilize resource estimation to compare the (potential) benefits gained by applying NISQ quantum circuit optimization to estimated FTQC resource requirements.
The results indicate that, indeed, the estimated resource requirements for FTQC can be improved by applying NISQ quantum circuit optimization techniques.
At the same time, more detailed investigations show what techniques lead to more benefits for FTQC compared to others, providing guidelines for the transfer of NISQ optimization techniques to the FTQC era.
\end{abstract}

\section{Introduction}
\label{sec:Introduction}
\vspace{3mm}
Quantum computing \cite{nielsen2010quantum} has made substantial strides in both hardware and software in the last few years.
This increasingly triggers interest in developing quantum applications in both academia and industry in various areas like \mbox{finance \cite{albareti2022structured}}, \mbox{chemistry \cite{santagati2024drug}}, and many more \cite{bayerstadler2021industry, ur2023quantum, flother2023state, 10313734, 10821384}.
Such applications are encoded into quantum circuits that comprise a sequence of quantum gates acting on qubits which are executed on a quantum computing device.
Yet, the hardware restrictions of the respective device (such as the native gate set or qubit topology) need to be taken into account. 
For this, quantum circuits are \emph{compiled} using three types of compilation passes, namely \emph{synthesis} \cite{furrutter2024quantum, di2016parallelizing, giles2013exact, miller2011elementary}, \emph{optimization} \cite{foesel2021quantum, liu2021relaxed, davis2020towards, younis2022quantum}, and \mbox{\emph{mapping} \cite{peham2023optimal, qmap, 10821450, hybrid, lin2023scalable, matsuo2019efficient, liu2023tackling}.}
This work focuses on the \emph{optimization} pass.

When translating the sequences of quantum gates into the native gate set of the chosen device (using a \mbox{\emph{synthesis}} pass), the number of gates potentially increases.
Since quantum hardware is heavily subject to errors, quantum circuits strongly suffer from noise and the more gates are executed, the worse it gets.
This behavior characterizes the \mbox{\emph{Noisy Intermediate-Scale Quantum} (NISQ)} era of quantum computing.
To counteract that, various \emph{optimization} techniques have been developed over the past decades with the intention of reducing the number of gates.

However, with increasing qubit and gate sizes, the noise introduced by hardware errors will dominate the \mbox{execution---inhibiting} the scaling of quantum applications to industrial-relevant problem sizes.
To overcome this issue, various \mbox{error-correction} techniques (such as, \mbox{e.g., \cite{gottesman1997stabilizer, Kitaev_1997, steane1999efficient}}) and corresponding methods and tools (such as, \mbox{e.g., \cite{PRXQuantum, Berent2024decodingquantum}}) have been developed  aiming at detecting and correcting \mbox{errors---shifting} quantum computing from the NISQ era to the \mbox{\emph{Fault-Tolerant Quantum Computing} (FTQC)} era. 
By this, qubit overheads are introduced (often of orders of magnitude), leading to tens of thousands of qubits already for toy-sized problems.

Obviously, these magnitudes are unrealistic, and, hence, quantum software developers start to focus on developing tools for quantum circuit \emph{optimization} that are also applicable for the FTQC era.
Since the \emph{optimization} techniques mentioned above are capable of improving the number of gates for NISQ quantum circuits, this raises the question, whether we can simply re-use these techniques as a basis for the FTQC era or whether we need to start from scratch. 

In this work, we approach this question by investigating the effects of different \emph{optimization} \mbox{passes---implemented} as part of IBM's quantum \emph{Software Development Kit} (SDK) \mbox{Qiskit \cite{qiskit2024}} and Quantinuum's SDK \mbox{TKET~\cite{tket}---on} a representative selection of quantum circuits, extracted from MQT Bench \cite{quetschlich2023mqtbench}.
As hardly any tools are available yet that convert NISQ quantum circuits to an FTQC version, \mbox{resource estimation \cite{10.1145/3624062.3624211,beverland2022assessingrequirementsscalepractical, 10821277}} is utilized to evaluate the FTQC resource requirements.
For this, the quantum circuits undergo a \emph{synthesis} to a gate set that allows a straightforward evaluation, an estimation of the FTQC resource requirements, an \emph{optimization}, and another estimation.
Hereby, benefits for the NISQ era are compared to benefits for the FTQC era when using currently available \emph{optimization} passes.

The results of these investigations confirm that the optimization tools developed in the past years for NISQ circuits can improve FTQC hardware resource requirements. 
That is, FTQC quantum circuit optimization does \emph{not} have to be developed from scratch.
At the same time, the investigations provide indications of what optimization effects lead to stronger improvements for FTQC compared to others.
Overall, providing these insights provides a foundation for the transfer of NISQ optimization techniques to the FTQC era, where different metrics are crucial for successful optimization.

The remainder of this work is structured as follows. 
\autoref{sec:Background} reviews the basics of quantum computing in general as well as \mbox{fault-tolerant} quantum computing.
\autoref{sec:Motivation} briefly summarizes the basics of quantum circuit compilation and \emph{optimization} as well as discusses obstacles for the transition from the NISQ to the FTQC era.
The tool used to estimate FTQC resources is described in \autoref{sec:Resource Estimation} before \autoref{sec:Evaluation} provides a comprehensive overview of the performance of quantum circuit \emph{optimizers} for the NISQ and FTQC era.

\section{Background}
\label{sec:Background}

\subsection{Quantum Computing}
\label{subsec:Quantum Computing}

Quantum bits extend the functionality of classical bits by not limiting their state to being either $0$ or $1$, but a superposition of these as well.
Hence, the state of a qubit $\ket{\psi}$ is written in the form of amplitudes of the basic quantum states $\ket{0}$ and $\ket{1}$, as
$\ket{\psi} = \alpha_0 \ket{0} + \alpha_1 \ket{1}$ with $\alpha_0$, $\alpha_1 \in \mathbb{C}$ and $|\alpha_0|^2 + |\alpha_1|^2 =1$.
In order to identify a qubit's state, a measurement must be performed that inevitably destroys a superposition state and forces a collapse to one of its basis states.
The probabilities with which the qubit collapses to a basic state are given by their respective squared amplitude as $|\alpha_i|^2$.

\begin{example}
\label{ex:quantum state}
    If a qubit is in the superposition state \mbox{$\ket{\psi} = 1/\sqrt{2} \ket{0} + 1/\sqrt{2} \ket{1}$}, a measurement would collapse the superposition and the qubit would result in either \mbox{$\ket{0}$ or $\ket{1}$}. 
    The probabilities for each state are equal in this case and given by $|1/\sqrt{2}|^2 = 0.5$.
\end{example}

Mathematically, the state of one or multiple qubits can also be expressed as a \emph{statevector}, whose entries represent the amplitudes of the basis states. 

\begin{example}
    The state \mbox{$\ket{\psi} = \alpha_0 \ket{0} + \alpha_1 \ket{1}$} can also be written as
    $
    \begin{bmatrix}
    \alpha_0 \:	
    \alpha_1 
  \end{bmatrix}^T$.
    When considering a system of two qubits, the statevector extends to four entries representing the amplitudes of the basis states $\ket{00}$, $\ket{01}$, $\ket{10}$, and $\ket{11}$.
\end{example}

Correspondingly to classical computers, quantum computers use quantum gates to perform logical operations on qubits and transform their state.
These quantum gates can either act on a single or on multiple qubits simultaneously. 
A key representative of a single-qubit quantum gate is the \emph{Hadamard} ($H$) gate. 
It exploits the quantum nature of qubits by transforming a qubit in state $\ket{0}$ or $\ket{1}$ to a superposition state.
An example of a two-qubit quantum gate is the \emph{CNOT} gate. 
It performs a switch of amplitudes of the state of one qubit (the \emph{target qubit}), if the state of another qubit (the \emph{control qubit}) equals $\ket{1}$.
If the state of the \emph{control qubit} equals $\ket{0}$, no transformation is applied to the \emph{target qubit}.

Overall, there exists a broad spectrum of different quantum \mbox{gates---like} Pauli or rotation \mbox{gates---that} transform a qubit's state in various different ways.
An overview of different quantum gates as well as a comparison to classical counterparts is provided by \cite{williams2011quantum}.
When arranged in a sequence to encode and process information for a given application, quantum gates form a \mbox{so-called} \emph{quantum circuit}.

\begin{figure*}[t]
\centering
\resizebox{0.99\linewidth}{!}{
\begin{tikzpicture}
\begin{yquantgroup}
\registers{
qubit {} q[2];
}
\circuit{
        init {$q_0$} q[0];
        init {$q_1$} q[1];
        h q[1] | q[0];
}
\equals[$\Longrightarrow$]
\circuit{
        box {$S$} q[1];
        box {$H$} q[1];
		box {$T$} q[1];   
		cnot q[0] | q[1];
		box {$R_Z(\frac{-\pi}{4})$} q[1];
		box {$H$} q[1];
		box {$R_Z(\frac{-\pi}{2})$} q[1];
}
\equals[$\Longrightarrow$]
\circuit{
        box {$R_X(\frac{-\pi}{2})$} q[1];
        box {$R_Z(\frac{-\pi}{4})$} q[1];
        cnot q[0] | q[1];
        box {$R_Z(\frac{\pi}{4})$} q[1];
        box {$R_X(\frac{\pi}{2})$} q[1];
}
\end{yquantgroup}
\end{tikzpicture}
}
\subfloat[Quantum circuit.\label{fig:original_qc} ]{\hspace{.15\linewidth}}
\hspace{0mm}
\subfloat[Synthesized circuit.\label{fig:syn_qc}]{\hspace{.4\linewidth}}
\hspace{0mm}
\subfloat[Optimized circuit.\label{fig:opt_qc}]{\hspace{.43\linewidth}}
\caption{Quantum circuit synthesis and optimization.}
\vspace{-2mm}
\label{fig:compilation}
\end{figure*}
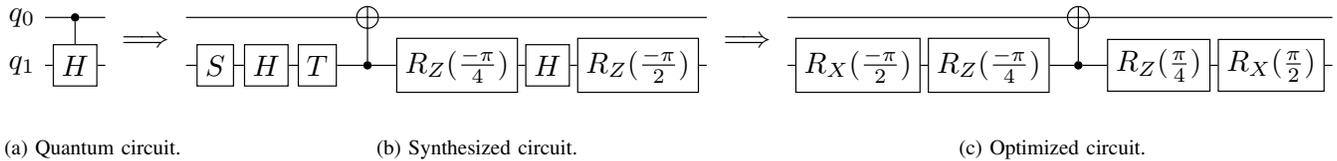

\subsection{Fault-Tolerant Quantum Computing}
\label{subsec:Fault Tolerant Quantum Computing}

Quantum circuits can be executed by both simulators and current NISQ devices. Unfortunately, simulators are very limited by their number of simulatable qubits while the NISQ devices suffer from severe execution \mbox{errors---getting} more severe with an increasing number of qubits and gates. Therefore, to execute larger circuits, quantum computing needs to include \mbox{\mbox{error-correction}} to counteract this behavior and execute quantum algorithms in a \mbox{fault-tolerant} fashion.
This section briefly summarizes the concepts of quantum \mbox{error-correction} that can be studied in more detail in \cite{roffe2019quantum}.

As described in \autoref{subsec:Quantum Computing}, the quantum state of a qubit is continuous due to the continuous amplitudes $\alpha_i$.
Therefore, the errors in quantum computing must be considered continuous as \mbox{well---exceeding} the classical bit-flip errors.
Besides that, the no-cloning theorem \cite{wootters1982single} states the impossibility of cloning quantum states and, hence, cloning a qubit's state to another.
Thus, to account for the destruction of quantum states during a measurement, continuous quantum errors and the no-cloning theorem, quantum \mbox{error-correction} techniques are significantly more complex than classical \mbox{error-correction} techniques.
That is, classical \mbox{error-correction} techniques cannot simply be transferred to quantum computing.
Instead, quantum \mbox{error-correction} is a comprehensive field of research to overcome the challenges mentioned above.

With the aim of detecting and correcting quantum errors, quantum \mbox{error-correction} techniques (or \mbox{error-correction} \emph{codes}) use redundant qubits in a very complex fashion. 
To this end, the information of one \emph{logical qubit} is encoded into many \emph{physical qubits} in a way that respects the no-cloning theorem.
Unfortunately, this introduces significant qubit overheads, due to the complexity of these processes, often in the orders of magnitude.
The exact way and procedure of encoding qubits as well as detecting and correcting errors depends on the used \mbox{error-correction} \emph{code} (e.g. the \mbox{surface code \cite{fowler2012surface, litinski2019game, google2023suppressing}}) and decoding algorithm.
The amount of \emph{physical qubits} needed to encode a \emph{logical qubit} depends, besides other factors, on the type of error correction \emph{code} and the relation of error rates of the \emph{physical qubits} to allowed logical error rates overall.
Despite the variety of currently researched \mbox{error-correction} approaches, they all lead to tens of thousands of \emph{physical qubits} even for quantum circuits comprised of just a handful of \emph{logical qubits}.

\section{Motivation}
\label{sec:Motivation}
This section briefly reviews quantum circuit compilation, especially optimization, as it is currently applied for NISQ quantum circuits.
Afterwards, we review the (upcoming) challenges when optimizing FTQC quantum \mbox{circuits---motivating} this work.
\subsection{Quantum Circuit Compilation and Optimization}
\label{subsec:Quantum Circuit Compilation and Optimization}

Quantum circuits are comprised of a sequence of instructions, so called gates, that manipulate its qubits. 
When aiming at executing a quantum circuit on a device, its hardware restrictions (such as the native gate set or qubit topology) need to be considered.
To this end, the quantum circuit is \emph{compiled} to adhere to the restrictions of the \mbox{device---while} keeping its functionality equal. 
Such a compilation procedure usually consists of three passes: \mbox{\emph{synthesis}, \emph{optimization}, and \emph{mapping}}. 
In the following, we will focus on the \emph{optimization} passes.
The \emph{synthesis} pass translates the set of all gates of a circuit to a sequence of gates actually executable on the chosen device. 
\begin{example}
	Consider the quantum circuit given in \autoref{fig:original_qc}, and a device with a native gate set of \{S, H, T, CX, RZ\}.
	Since the CH gate is not an element of the native gate set, it must be synthesized to a sequence of native gates realizing CH. 
	This results in the circuit as shown \autoref{fig:syn_qc}.
\end{example}
Since the translated circuit potentially consists of more gates than the original circuit, \emph{optimization} passes are used to reduce this gate overhead. 
\begin{example}
	By replacing some gates of the circuit with other native \mbox{gates---e.g.,} the S, H and T gate with an RX and RZ \mbox{gate---the} number of gates (or gate count) of the circuit can be reduced.
	This results in the circuit as shown in \autoref{fig:opt_qc}.
\end{example}
Depending on the structure of the circuit and the native gate set, the effectiveness of \emph{optimizations} can vary significantly. 
An \emph{optimization} pass can potentially change the set of gates again, such that parts of the circuit are not in the native gate set anymore. 
It can therefore take multiple iterations of \emph{synthesis} and \emph{optimization} passes to yield an \emph{optimized} circuit consisting of native gates only. 

Due to the importance of quantum circuit compilation, many techniques for \mbox{\emph{synthesis} \cite{furrutter2024quantum, di2016parallelizing, giles2013exact, miller2011elementary}}, \mbox{\emph{optimization} \cite{foesel2021quantum, liu2021relaxed, davis2020towards, younis2022quantum}}, and mapping~\cite{peham2023optimal, qmap, 10821450, hybrid, lin2023scalable, matsuo2019efficient, liu2023tackling} have been developed. 
Together with many additional other functionalities and methods, these techniques are orchestrated by \emph{compilers} usually embedded in quantum \mbox{\emph{Software Development Kits} (SDKs)}.
These realize automated and easily usable but very complex compilation flows. 
Both industry and academia offer a variety of such SDKs. 
Some representatives for the industry are IBM's Qiskit \cite{qiskit2024}, Quantinuum’s TKET \cite{tket}, and Xanadu's PennyLane \cite{bergholm2022pennylane}. 
But also academic SDKs become more prominent and comprehensive, e.g., the Berkeley Quantum Synthesis Toolkit \cite{osti_1785933} and the Munich Quantum Toolkit \cite{willeMQTHandbookSummary2024}.

\vspace{-2mm}
\subsection{Optimization in FTQC}
\label{subsec:Optimization in FTQC}

The flow reviewed above focuses on optimizing circuits of the \emph{Noisy Intermediate-Scale Quantum} (NISQ) era. 
However, current quantum hardware suffers from noise introduced by error-prone qubits.
This becomes particularly problematic with increasing problem sizes 
due to the increased number of qubits and gates.
In the worst case, this can lead to a situation where the probability with which a quantum circuit execution yields the desired result considering certain error rates of the qubits (or \emph{expected fidelity}) decreases to an unusable level.

\begin{example}
	Consider an average gate fidelity of $99.9 \%$ (which is already very optimistic) and a quantum circuit comprised of $1,000$ gates. 
	This would yield an expected fidelity of the overall circuit of $\sim 36.8 \%$. 
	When increasing the gate count to $5,000$, the expected fidelity would decrease to \mbox{$\sim 0.7 \%$---and,} therefore, leading to almost meaningless results.
\end{example}
To overcome this issue and suppress quantum errors, executions of quantum circuits need to incorporate \emph{\mbox{error-correction}} techniques (such as, e.g.,~\cite{gottesman1997stabilizer, Kitaev_1997, steane1999efficient}).
For this, various \mbox{\emph{error-correction}} codes and methods (such as, e.g.,~\cite{PRXQuantum, Berent2024decodingquantum}) have been developed that encode information of one \emph{logical qubit} into many \emph{physical qubits} and execute operations in a \mbox{fault-tolerant} fashion.
A \emph{logical qubit} is an abstract object described in a quantum circuit, whereas a \emph{physical qubit} is an actual qubit of a physical quantum device.
This redundancy is used to detect and correct errors before the information is decoded \mbox{again---leading} to the \mbox{\emph{Fault-Tolerant Quantum Computing} (FTQC)} era.
However, at the same time, this dramatically induces qubit \mbox{overheads---often} of orders of magnitude. 
\begin{example}
	To execute the quantum circuit from \autoref{fig:syn_qc} on a device with a gate fidelity of $99.9 \%$ in a \mbox{fault-tolerant} fashion, approximately $5,000$ physical qubits would be required to keep the logical error rate at a maximum of $0.1 \%$. 
	This assumes a naive \mbox{fault-tolerant} implementation of the circuit without any optimization concerning the required physical qubits.
\end{example}

Obviously, executing a toy example as above with thousands of qubits is unrealistic.
Hence, quantum software developers start to focus on optimizing quantum circuits with respect to the \mbox{fault-tolerant} hardware requirements to accelerate the transition from the NISQ era to the FTQC era.
This constitutes a very challenging task.
And while quantum circuit optimization has already been investigated for several years (leading to the flow and methods briefly reviewed in \autoref{subsec:Quantum Circuit Compilation and Optimization}), hardly any tools are currently available for quantum circuit optimization in the FTQC era.
This raises the \mbox{question: Do we,} for this purpose, have to start from scratch, or can we re-use the state of the art optimizers from the NISQ era as a basis?

In this work, we address this question by investigating the effects of different optimization passes on a representative selection of quantum circuits.
Due to a lack of suitable tools for designing and evaluating FTQC quantum circuits, we utilize resource estimation for this purpose. 
In the remainder of this work, the corresponding approach and the results including the obtained findings are described.
More precisely, \autoref{sec:Resource Estimation} introduces the implementation of the investigation which utilizes Microsoft's Azure Quantum \mbox{Resource Estimator~\cite{10.1145/3624062.3624211,beverland2022assessingrequirementsscalepractical}} allowing to evaluate FTQC benefits of circuits generated with current NISQ optimization tools in the absence of FTQC design tools.
Based on that, \autoref{sec:Evaluation} then presents the obtained findings, provides detailed insights, and draws conclusions.
Overall, these findings aim to provide a foundation for quantum circuit optimization in the FTQC \mbox{era---a} crucial and challenging design step which is essential for the success of quantum computing in the coming era.

\begin{table*}[t]
    \centering
    \caption{Relative changes in percent for the considered Qiskit optimization passes.}  %
    \resizebox{0.99\textwidth}{!}{
    \begin{tabular}{l?R|RR?R|RR?R|RR?R|RR?R|RR}
    \multicolumn{1}{c?}{} & \multicolumn{3}{c?}{Commutative-} & \multicolumn{3}{c?}{Optimize-} & \multicolumn{3}{c?}{Optimize-} & \multicolumn{3}{c?}{Hoare-} & \multicolumn{3}{c}{Template-}  \\
    
    \multicolumn{1}{c?}{} & \multicolumn{3}{c?}{Cancellation} & \multicolumn{3}{c?}{1qGatesDecomposition} & \multicolumn{3}{c?}{1qGatesSimpleCommutation} & \multicolumn{3}{c?}{Optimizer} & \multicolumn{3}{c}{Optimization}  \\ \hline
    
    \multicolumn{1}{c?}{} & \multicolumn{1}{c|}{NISQ} & \multicolumn{2}{c?}{FTQC} & \multicolumn{1}{c|}{NISQ} & \multicolumn{2}{c?}{FTQC} & \multicolumn{1}{c|}{NISQ} & \multicolumn{2}{c?}{FTQC} & \multicolumn{1}{c|}{NISQ} & \multicolumn{2}{c?}{FTQC} & \multicolumn{1}{c|}{NISQ} & \multicolumn{2}{c}{FTQC} \\ \hline

    \multicolumn{1}{c?}{} & \multicolumn{1}{c|}{\#G} & \multicolumn{1}{c}{\#Q} & \multicolumn{1}{c?}{t} & \multicolumn{1}{c|}{\#G} & \multicolumn{1}{c}{\#Q} & \multicolumn{1}{c?}{t} & \multicolumn{1}{c|}{\#G} & \multicolumn{1}{c}{\#Q} & \multicolumn{1}{c?}{t} & \multicolumn{1}{c|}{\#G} & \multicolumn{1}{c}{\#Q} & \multicolumn{1}{c?}{t} & \multicolumn{1}{c|}{\#G} & \multicolumn{1}{c}{\#Q} & \multicolumn{1}{c}{t} \\ \hline
    
    ae & \cellcolorval{-11} & \cellcolorval{-44} & \cellcolorval{-15} & \cellcolorval{-25} & \cellcolorval{0} & \cellcolorval{0} & \cellcolorval{-25} & \cellcolorval{0} & \cellcolorval{0} & \cellcolorval{-3} & \cellcolorval{0} & \cellcolorval{0}& \cellcolorval{0} & \cellcolorval{+11} & \cellcolorval{-11} \\ \hline

    dj & \cellcolorval{-9} & \cellcolorval{0} & \cellcolorval{0} & \cellcolorval{-41} & \cellcolorval{0} & \cellcolorval{0} & \cellcolorval{-41} & \cellcolorval{0} & \cellcolorval{0} & \cellcolorval{-11} & \cellcolorval{0} & \cellcolorval{0} & \cellcolorval{0} & \cellcolorval{0} & \cellcolorval{0} \\ \hline

    graphstate & \cellcolorval{-39} & \cellcolorval{0} & \cellcolorval{0} & \cellcolorval{-39} & \cellcolorval{0} & \cellcolorval{0} & \cellcolorval{-31} & \cellcolorval{0} & \cellcolorval{0} & \cellcolorval{-27} & \cellcolorval{0} & \cellcolorval{0} & \cellcolorval{0} & \cellcolorval{0} & \cellcolorval{0} \\ \hline

    grover-noancilla & \cellcolorval{-2} & \cellcolorval{0} & \cellcolorval{0} & \cellcolorval{-4} & \cellcolorval{0} & \cellcolorval{0} & \cellcolorval{-4} & \cellcolorval{0} & \cellcolorval{0} & \cellcolorval{-10} & \cellcolorval{-22} & \cellcolorval{-2} & \cellcolorval{0} & \cellcolorval{0} & \cellcolorval{0} \\ \hline

    grover-v-chain & \cellcolorval{-5} & \cellcolorval{0} & \cellcolorval{0} & \cellcolorval{-8} & \cellcolorval{0} & \cellcolorval{0} & \cellcolorval{-8} & \cellcolorval{0} & \cellcolorval{0} & \cellcolorval{-4} & \cellcolorval{-41} & \cellcolorval{-1} & \cellcolorval{0} & \cellcolorval{0} & \cellcolorval{0} \\ \hline
    
    qft & \cellcolorval{-14} & \cellcolorval{-40} & \cellcolorval{-23} & \cellcolorval{0} & \cellcolorval{0} & \cellcolorval{0} & \cellcolorval{0} & \cellcolorval{0} & \cellcolorval{0} & \cellcolorval{-17} & \cellcolorval{-43} & \cellcolorval{-27} & \cellcolorval{0} & \cellcolorval{-24} & \cellcolorval{+34} \\ \hline

    qftentangled & \cellcolorval{-13} & \cellcolorval{-40} & \cellcolorval{-23} & \cellcolorval{0} & \cellcolorval{0} & \cellcolorval{0} & \cellcolorval{0} & \cellcolorval{0} & \cellcolorval{0} & \cellcolorval{0} & \cellcolorval{0} & \cellcolorval{0} & \cellcolorval{0} & \cellcolorval{-24} & \cellcolorval{+34} \\ \hline
    
    qpeexact & \cellcolorval{-14} & \cellcolorval{-38} & \cellcolorval{-28} & \cellcolorval{0} & \cellcolorval{0} & \cellcolorval{-2} & \cellcolorval{0} & \cellcolorval{0} & \cellcolorval{-2} & \cellcolorval{-3} & \cellcolorval{+21} & \cellcolorval{-22} & \cellcolorval{0} & \cellcolorval{-3} & \cellcolorval{+2} \\ \hline

    qpeinexact & \cellcolorval{-14} & \cellcolorval{-38} & \cellcolorval{-28} & \cellcolorval{0} & \cellcolorval{0} & \cellcolorval{0} & \cellcolorval{0} & \cellcolorval{0} & \cellcolorval{0} & \cellcolorval{-3} & \cellcolorval{+32} & \cellcolorval{-30} & \cellcolorval{0} & \cellcolorval{-4} & \cellcolorval{+4} \\ \hline

    qwalk-noancilla & \cellcolorval{-2} & \cellcolorval{+5} & \cellcolorval{-1} & \cellcolorval{-2} & \cellcolorval{0} & \cellcolorval{0} & \cellcolorval{-2} & \cellcolorval{0} & \cellcolorval{0} & \cellcolorval{-9} & \cellcolorval{+5} & \cellcolorval{-8} & \cellcolorval{0} & \cellcolorval{0} & \cellcolorval{0} \\ \hline

    qwalk-v-chain & \cellcolorval{-11} & \cellcolorval{0} & \cellcolorval{-4} & \cellcolorval{-14} & \cellcolorval{+7} & \cellcolorval{-12} & \cellcolorval{-14} & \cellcolorval{+7} & \cellcolorval{-12} & \cellcolorval{-31} & \cellcolorval{+15} & \cellcolorval{-45} & \cellcolorval{-2} & \cellcolorval{0} & \cellcolorval{0} \\ \hline

    \end{tabular} 
    }
    \label{tab:results_qiskit}
    \vspace{-3mm}
\end{table*}

\section{Resource Estimation for FTQC}
\label{sec:Resource Estimation}

Due to the dramatic qubit overheads introduced by FTQC and reviewed in \autoref{subsec:Fault Tolerant Quantum Computing}, executing or simulating error-corrected quantum circuits becomes challenging.
When including \mbox{error-correction} into the development of a quantum application, the resulting quantum circuits become significantly more complex.
Currently, hardly any tools are available yet to design and evaluate quantum circuits for FTQC.
Thus, \emph{resource estimation} is an evolving domain that is used to estimate the required hardware resources for a quantum application when utilizing \mbox{error-correction} instead of \mbox{executing it}.
Despite not providing FTQC solutions or quantum circuits, it provides information about the FTQC resources required, that can be integrated into the development of a quantum application~\cite{10821277}.
Prominent examples of \emph{resource estimation} tools are Microsoft's Azure Quantum Resource Estimator~\cite{10.1145/3624062.3624211, beverland2022assessingrequirementsscalepractical}, Google's Qualtran~\cite{harrigan2024expressinganalyzingquantumalgorithms}, and MIT Lincoln Lab's pyLIQTR~\cite{obenland_2025_14719561}.

In this work, we utilize Microsoft's Azure Quantum Resource Estimator, as described in \cite{10.1145/3624062.3624211}. 
This section first reviews the tool.
Afterward, we describe in detail how we used that tool for the purposes of this work.

\subsection{Microsoft Azure Quantum Resource Estimator}
\label{subsec:MSRE}

Microsoft's Azure Quantum Resource Estimator takes a quantum circuit as its input based on a format that is comprised of seven categories: the number of logical qubits, T gates, single-qubit rotation gates, rotation gate layers, CCZ gates, CCiX gates, and measurements.
By utilizing a layout that allows for eliminating Clifford gates (such as X, CNOT or H gates) with the cost of performing multi-qubit Pauli measurements that do not consume additional resources, Clifford gates are ignored by the tool.

Alongside with these inputs, the resource estimator offers a spectrum of customization to adjust the assumed underlying hardware and \mbox{error-correction} characteristics like the qubit model, \mbox{error-correction} code properties, allowed logical error budget, and many more.
Based on the seven input parameters characterizing a quantum circuit and the customized parameters, the tool provides the end-user with an estimation of the required runtime and physical qubits.
Additionally, through an assumption of different hardware configurations, the resource estimator provides a trade-off between the estimated runtime and physical qubits.
That is, the end-user can choose between multiple configurations that prefer shorter runtimes at the cost of more physical qubits and vice versa.

\subsection{Proposed Approach}
\label{subsec:Implementation}

Since this work aims to investigate the effects of various optimization passes on FTQC hardware requirements and compare these effects to the NISQ metric gate count, certain steps are conducted initially to ensure a consistent and fair comparison.

To this end, the quantum circuits are translated to a gate set consisting of only single-qubit gates plus the CNOT gate.
Additionally, gates that compose multiple gates at once (the U gate in Qiskit and the TK gates in TKET) are eliminated to ensure a consistent gate count.
Otherwise, they would be counted as one gate but would actually need multiple gates to realize.
The resulting target gate set consists of X, Y, Z, H, S, Sdg, T, Tdg, SX, RX, RY, RZ, CNOT gates.
To enable using the estimator for the TKET circuits, they are always converted to a Qiskit circuit beforehand.

Then, the gate counts of the resulting circuits are determined, and an estimation of the required FTQC resources is conducted using the resource estimator.
Subsequently, the considered optimization pass is applied to the circuits. 
Since some optimization passes can change the gate set of a quantum circuit, the same translation step as above is applied again after the optimization.

In the case of TKET, this includes a custom decomposition of the TK1 gate into either various Clifford gates or a combination of rotation gates.
Otherwise, TKET's optimization passes would potentially compose all gates in the circuits to CNOT and TK1 gates, whereby the latter ones would be decomposed to three rotations in the subsequent step of translating back to the desired gate set.
However, it is not always required to utilize three rotations for this decomposition, but there exist many special cases in which, e.g., only one Clifford gate is sufficient.
Again, the gate count of the resulting circuits is determined, and the FTQC resources are estimated.

\section{Obtained Findings}
\label{sec:Evaluation}

This section provides an overview of the findings \mbox{obtained} by applying various NISQ optimization techniques to a selection of representative quantum circuits.
Based on this overview, we approach the question of whether the development of quantum circuit optimization techniques needs to be started from scratch or if NISQ optimization techniques can serve as a basis for the optimization of FTQC.
This investigation aims to guide quantum software \mbox{developers---particularly} NISQ \mbox{developers---on} what is important when optimizing quantum circuits for the transition to the era of FTQC.
By extracting the underlying effects leading to the obtained results, this section provides a set of aspects that should be considered when \mbox{optimizing quantum circuits with respect to FTQC}.

\subsection{Evaluation Setup}
\label{subsec:Setup}

The considered quantum circuits were generated with MQT Bench~\cite{quetschlich2023mqtbench} as 10-qubit circuits (7-qubit for the Grover and qwalk circuits and 5-qubit for qwalk-v-chain with HoareOptimizer) in two versions, each for Qiskit v1.4.2 and TKET v2.1.0.
To conduct the resource estimation, \mbox{Microsoft's Azure Quantum Resource Estimator} was used within the "qsharp" v1.14.0 Python package using its default parameters and the implementation is available on GitHub (\url{https://github.com/cda-tum/mqt-problemsolver}) as part of the \emph{Munich Quantum Toolkit}~\cite{willeMQTHandbookSummary2024}.
The considered optimization passes by Qiskit were
\begin{itemize}
    \item CommutativeCancellation,
    \item Optimize1qGatesDecomposition,
    \item Optimize1qGatesSimpleCommutation,
    \item HoareOptimizer, and
    \item TemplateOptimization.
\end{itemize}
Regarding TKET, the passes were 
\begin{itemize}
    \item CliffordSimp,
    \item FullPeepholeOptimise,
    \item RemoveRedundancies,
    \item KAKDecomposition, and
    \item CommuteThroughMultis + RemoveRedundancies.
\end{itemize}

The influences of the considered optimization passes of Qiskit and TKET are summarized in \autoref{tab:results_qiskit} and \autoref{tab:results_tket}, respectively.
The results are split into NISQ and FTQC metrics using the gate count for NISQ (\#G) and the estimated physical qubits (\#Q) as well as the estimated runtime (t) for FTQC.
The numbers refer to relative differences of the metrics between the originally given circuits and the optimized ones.
Negative numbers (-) indicate a \mbox{reduction---positive} numbers (+) an increase.
These numbers led to three concrete findings, which are discussed next.

\subsection{Finding 1: Optimizing Clifford Gates Improves NISQ More Than FTQC}
\label{subsec:Finding 1}

It is observed that optimizing Clifford gates by merging multiple ones or removing redundant sequences thereof is primarily beneficial for the NISQ metric gate count. 
This is shown in cases in which the NISQ column is negative, denoted by the green color, while both FTQC columns are indifferent, indicated by no color, showing no changes.
Examples of this behavior are the Deutsch-Jozsa (\enquote{dj}) algorithm when being optimized with Qiskit's \enquote{Optimize1qGatesDecomposition} or Amplitude Estimation (\enquote{ae}) during TKET's \enquote{CliffordSimp} with a reduction of 41\% and 9\% in gate count and no differences in the FTQC metrics. 

In FTQC, a particular computation scheme (Parallel Synthesis Sequential Pauli Computation, PSSPC~\cite{beverland2022assessingrequirementsscalepractical}) that allows for a highly resource-efficient implementation of Clifford gates can be applied.
With this, the additional costs of implementing Clifford gates compared to other gates is negligible, which is why the resource estimator does not take them into account.
Hence, even though the gate count can potentially be significantly reduced, eliminating Clifford gates does not improve the estimated FTQC resource requirements.
This also holds for rotation gates with rotation angles that are multiples of $\frac{\pi}{2}$ as they can be replaced by Clifford gates without changing their functionality.
The cases of increased gate counts in the frame of TKET occur due to the imperfect decompositions of TK1 gates into the desired gate set after the optimization.
That is, during optimization, many gates are combined into a TK1 gate that needs to be decomposed subsequently into gates of the desired set.
In some cases, this can lead to an increased gate count as, e.g., a single gate gets transformed into a TK1 gate, followed by a decomposition into multiple gates, as some TK1 gates are not decomposed the most efficient way. 

\begin{table*}[t]
    \centering
    \caption{Relative changes in percent for the considered TKET optimization passes.}  %
    \resizebox{0.99\textwidth}{!}{
    \begin{tabular}{l?R|RR?R|RR?R|RR?R|RR?R|RR}
    \multicolumn{1}{c?}{} & \multicolumn{3}{c?}{Clifford-} & \multicolumn{3}{c?}{Full-} & \multicolumn{3}{c?}{Remove-} & \multicolumn{3}{c?}{KAK-} & \multicolumn{3}{c}{CommuteThroughMultis +}  \\
    
    \multicolumn{1}{c?}{} & \multicolumn{3}{c?}{Simp} & \multicolumn{3}{c?}{PeepholeOptimise} & \multicolumn{3}{c?}{Redundancies} & \multicolumn{3}{c?}{Decomposition} & \multicolumn{3}{c}{RemoveRedundancies}  \\ \hline

    \multicolumn{1}{c?}{} & \multicolumn{1}{c|}{NISQ} & \multicolumn{2}{c?}{FTQC} & \multicolumn{1}{c|}{NISQ} & \multicolumn{2}{c?}{FTQC} & \multicolumn{1}{c|}{NISQ} & \multicolumn{2}{c?}{FTQC} & \multicolumn{1}{c|}{NISQ} & \multicolumn{2}{c?}{FTQC} & \multicolumn{1}{c|}{NISQ} & \multicolumn{2}{c}{FTQC} \\ \hline
    
    \multicolumn{1}{c?}{} & \multicolumn{1}{c|}{\#G} & \multicolumn{1}{c}{\#Q} & \multicolumn{1}{c?}{t} & \multicolumn{1}{c|}{\#G} & \multicolumn{1}{c}{\#Q} & \multicolumn{1}{c?}{t} & \multicolumn{1}{c|}{\#G} & \multicolumn{1}{c}{\#Q} & \multicolumn{1}{c?}{t} & \multicolumn{1}{c|}{\#G} & \multicolumn{1}{c}{\#Q} & \multicolumn{1}{c?}{t} & \multicolumn{1}{c|}{\#G} & \multicolumn{1}{c}{\#Q} & \multicolumn{1}{c}{t} \\ \hline
    
    ae & \cellcolorval{-9} & \cellcolorval{0} & \cellcolorval{0} & \cellcolorval{-19} & \cellcolorval{-44} & \cellcolorval{-15} & \cellcolorval{0} & \cellcolorval{0} & \cellcolorval{0} & \cellcolorval{0} & \cellcolorval{0} & \cellcolorval{0} & \cellcolorval{-10} & \cellcolorval{-44} & \cellcolorval{-15} \\ \hline
    
    dj & \cellcolorval{+10} & \cellcolorval{0} & \cellcolorval{0} & \cellcolorval{+10} & \cellcolorval{0} & \cellcolorval{0} & \cellcolorval{0} & \cellcolorval{0} & \cellcolorval{0} & \cellcolorval{0} & \cellcolorval{0} & \cellcolorval{0} & \cellcolorval{0} & \cellcolorval{0} & \cellcolorval{0} \\ \hline
    
    graphstate &  \cellcolorval{-39} & \cellcolorval{0} & \cellcolorval{0} & \cellcolorval{-39} & \cellcolorval{0} & \cellcolorval{0} & \cellcolorval{-39} & \cellcolorval{0} & \cellcolorval{0} & \cellcolorval{0} & \cellcolorval{0} & \cellcolorval{0} & \cellcolorval{-39} & \cellcolorval{0} & \cellcolorval{0} \\ \hline
    
    grover-noancilla & \cellcolorval{+30} & \cellcolorval{0} & \cellcolorval{0} & \cellcolorval{+30} & \cellcolorval{0} & \cellcolorval{0} & \cellcolorval{-1} & \cellcolorval{0} & \cellcolorval{-1} & \cellcolorval{0} & \cellcolorval{0} & \cellcolorval{0} & \cellcolorval{0} & \cellcolorval{0} & \cellcolorval{0} \\ \hline
    
    grover-v-chain & \cellcolorval{+14} & \cellcolorval{0} & \cellcolorval{+3} & \cellcolorval{+15} & \cellcolorval{0} & \cellcolorval{+4} & \cellcolorval{-3} & \cellcolorval{0} & \cellcolorval{0} & \cellcolorval{0} & \cellcolorval{0} & \cellcolorval{0} & \cellcolorval{-3} & \cellcolorval{0} & \cellcolorval{0} \\ \hline
    
    qft & \cellcolorval{-6} & \cellcolorval{-3} & \cellcolorval{-5} & \cellcolorval{-20} & \cellcolorval{-41} & \cellcolorval{-27} & \cellcolorval{0} & \cellcolorval{0} & \cellcolorval{0} & \cellcolorval{-6} & \cellcolorval{-3} & \cellcolorval{-5} & \cellcolorval{-14} & \cellcolorval{-3} & \cellcolorval{-22} \\ \hline
    
    qftentangled & \cellcolorval{-6} & \cellcolorval{-3} & \cellcolorval{-5} & \cellcolorval{-19} & \cellcolorval{-41} & \cellcolorval{-27} & \cellcolorval{0} & \cellcolorval{0} & \cellcolorval{0} & \cellcolorval{-6} & \cellcolorval{-3} & \cellcolorval{-5} & \cellcolorval{-13} & \cellcolorval{-3} & \cellcolorval{-22} \\ \hline
    
    qpeexact & \cellcolorval{-5} & \cellcolorval{0} & \cellcolorval{-3} & \cellcolorval{-18} & \cellcolorval{-40} & \cellcolorval{-28} & \cellcolorval{0} & \cellcolorval{0} & \cellcolorval{0} & \cellcolorval{-7} & \cellcolorval{0} & \cellcolorval{-3} & \cellcolorval{-14} & \cellcolorval{-38} & \cellcolorval{-28} \\ \hline
    
    qpeinexact & \cellcolorval{-4} & \cellcolorval{-4} & \cellcolorval{-5} & \cellcolorval{-18} & \cellcolorval{-38} & \cellcolorval{-33} & \cellcolorval{0} & \cellcolorval{0} & \cellcolorval{0} & \cellcolorval{-7} & \cellcolorval{-4} & \cellcolorval{-5} & \cellcolorval{-14} & \cellcolorval{-38} & \cellcolorval{-28} \\ \hline

    qwalk-noancilla & \cellcolorval{-3} & \cellcolorval{0} & \cellcolorval{0} & \cellcolorval{+12} & \cellcolorval{0} & \cellcolorval{-1} & \cellcolorval{-4} & \cellcolorval{0} & \cellcolorval{0} & \cellcolorval{-5} & \cellcolorval{0} & \cellcolorval{0} & \cellcolorval{-5} & \cellcolorval{0} & \cellcolorval{-1} \\ \hline

    qwalk-v-chain & \cellcolorval{-10} & \cellcolorval{+7} & \cellcolorval{-12} & \cellcolorval{-5} & \cellcolorval{0} & \cellcolorval{+13} & \cellcolorval{-14} & \cellcolorval{+7} & \cellcolorval{-12} & \cellcolorval{-13} & \cellcolorval{+7} & \cellcolorval{-12} & \cellcolorval{-14} & \cellcolorval{+7} & \cellcolorval{-12} \\ \hline
    
    \end{tabular} 
    }
    \label{tab:results_tket}
    \vspace{-3mm}
\end{table*}

\subsection{Finding 2: Reducing the Number of Rotation Gates Leads to Strong FTQC Benefits}
\label{subsec:Finding 2}

This finding concerns the cases where an optimization pass led to improvements in both NISQ and FTQC.
This effect is mainly induced by exploiting commutation rules, such that some rotation gates can be canceled or merged together, leading to fewer gates.
Obviously, this reduces the gate count, but additionally shows even stronger influences on the estimated FTQC resource requirements.
Examples of this behavior are the Quantum Fourier Transform (\enquote{qft}) or Quantum Phase Estimation (\enquote{qpe}) processed by Qiskit's \enquote{CommutativeCancellation} or TKET's \enquote{FullPeepholeOptimise}.
Here, even though the gate count is reduced during optimization (by \enquote{just} between 13\% and 20\%), the estimated FTQC resource requirements are reduced \mbox{significantly stronger (namely up to 41\%)}.

Since rotation gates are highly resource-intensive in the frame of FTQC, reducing their amounts leads to significant reductions in the estimated physical resource requirements.
Again, the angle of rotation is crucial since more resource-efficient gates can replace angles multiple of $\frac{\pi}{2}$. 
In contrast, rotation gates with other arbitrary angles must be approximated utilizing multiple T gates (or a single one for $\frac{\pi}{4}$ rotations), making them highly resource-intensive.
Since the resource estimator used for this evaluation is able to differentiate between these rotation angles and appropriately take them into account, the effects thereof are already included in the results.

\subsection{Finding 3: Rearranging the Circuit Leads to Redistributions of FTQC Metrics}
\label{subsec:Finding 3}

In the case of some optimization processes, the structure of the quantum circuit changes.
That is, some gates change their position by being commuted through the quantum circuit.
The types and numbers of gates are mostly not influenced by this process, such that the differences in the NISQ metric gate count are minor. 
In contrast, the FTQC metrics are significantly influenced in both positive and negative trends.
This can be considered as a trade-off between qubits and runtime since a positive trend in one always leads to a negative trend in the other.
An example of that is the Quantum Fourier Transform algorithm (\enquote{qft}) optimized with Qiskit's \enquote{TemplateOptimization} with no difference in the gate count but a reduction of 24\% in the estimated physical qubits for FTQC and an increase of 34\% in the estimated FTQC runtime.

The reason for this is that rearranging the circuit can influence the number of layers of rotation gates, hence changing the number of their parallel occurrences.
In FTQC, rotation gates with continuous rotation angles are approximated by a discrete sequence of Clifford and T gates.
Since the latter ones are highly resource-intensive to implement fault-tolerantly, rotation gates are as well, as they require multiple T gates for a single rotation gate.
That is, changing the arrangement of rotation gates in quantum circuits leads to, e.g., more physical qubits as more T gates are implemented in parallel.
In turn, this would reduce the required runtime.
Obviously, that is also possible vice versa, always trading off the required physical qubits and runtime.

\subsection{Summary: Do We Have to Start from Scratch?}
\label{subsec:Do We Have to}

The short answer is: No, we indeed can build upon existing NISQ optimization routines when developing such routines for FTQC.
But, there are some important caveats.
Most importantly, the types of gates that should be optimized differ.
While in NISQ, optimization routines aim for a \emph{general} gates reduction, in FTQC only the reduction of T and rotation gates (with angles $\neq \frac{\pi}{2}$ or multiples thereof) matters (as discussed in detail in \autoref{subsec:Finding 1} and \autoref{subsec:Finding 2}). 
Furthermore, FTQC offers a trade-off between the number of estimated physical qubits and the runtime.
However, it is not trivial to judge whether an improvement in one measure and a deterioration in the other is an overall gain (as discussed in detail in \autoref{subsec:Finding 3}). 
Therefore, existing NISQ optimization routines that lead to such a situation should \emph{not} be blindly applied but only if it suits the individual use-case.
Thus, the existing optimization routines should be fine-tuned for these \mbox{aspects---instead} of being entirely developed from scratch.

\vspace{1mm}
\section{Conclusion}
\label{sec:Conclusion}

Quantum computing hardware is heavily subject to errors, which becomes more problematic the more qubits and gates are used for an application.
Hence, quantum computing needs to incorporate \mbox{error-correction} and make the transition from the current NISQ to the FTQC era.
However, this introduces dramatic qubit overheads such that simulating or executing error-corrected quantum circuits becomes challenging.
Thus, optimization techniques that reduce the FTQC resource requirements are necessary as a fundamental step to design, simulate, and execute \mbox{error-corrected} quantum circuits.
This raises the question, whether current NISQ optimization techniques, that aim to reduce the NISQ gate counts, can be used as a basis for this task or if we have to \mbox{start from scratch}.

In this work, we found that we do not have to start from scratch since current optimization tools can serve as a basis, but they have some caveats.
Thus, the findings of this work should be incorporated into the development of FTQC optimization routines to specialize them for this era.
These findings were extracted from an investigation on the benefits of NISQ optimization passes on FTQC, indicating that NISQ optimization techniques can indeed yield benefits for the hardware requirements of error-corrected quantum circuits.
At the same time, the findings show, that not just a general reduction of gate count, but a focus on reducing rotation gates with unfavorable angles is essential for FTQC.
Additionally, in FTQC, the required physical qubits and runtime can be traded off, which some current optimization passes influence.
Since it is individual whether an improvement in one measure and a deterioration in the other is an overall benefit, existing NISQ optimization routines influencing this behavior should not be blindly applied but with consideration of the individual use-case.
However, this work should by no means discourage the development of entirely new optimization approaches specifically targeting FTQC, but rather highlights the potential of enhancing existing NISQ optimization techniques as a complementary approach.

\section*{Acknowledgments}

This work received funding from the European Research Council (ERC) under the European Union’s Horizon 2020 research and innovation program (grant agreement No. 101001318), was part of the Munich Quantum Valley, which is supported by the Bavarian state government with funds from the Hightech Agenda Bayern Plus, and has been supported by the BMK, BMDW, the State of Upper Austria in the frame of the COMET program, and the QuantumReady project within Quantum Austria (managed by the FFG).

\vspace{200 pt}

\renewcommand*{\bibfont}{\fontsize{9}{11}\selectfont}
\printbibliography

\end{document}